\documentstyle[pra,aps,twocolumn]{revtex}

\input epsf

\begin{document}

\title{Modulational instability of solitary waves in 
non-degenerate three-wave mixing:\newline The role of phase symmetries.}
\author{Dmitry V. Skryabin 
\footnote{URL: http://cnqo.phys.strath.ac.uk/$\sim$dmitry}, 
and William J. Firth}
\address{Department of
Physics and Applied Physics, John Anderson Building,\\ 
University
of Strathclyde, 107 Rottenrow, Glasgow, G4 0NG, Scotland}

\date{April 8, 1998}

\maketitle

\begin{abstract}
We show how the analytical approach of Zakharov and Rubenchik 
[Sov. Phys. JETP {\bf 38}, 494 (1974)] 
to modulational instability (MI) of solitary waves 
in the nonlinear Schr\"oedinger equation (NLS) can be
generalised for models with  two phase symmetries.  
MI of three-wave parametric spatial 
solitons  due to group velocity dispersion (GVD) is investigated 
as a typical example of such models. We reveal a new 
branch of neck instability, which dominates the usual snake type MI 
found for normal GVD. The resultant nonlinear evolution is thereby 
qualitatively different from cases with only a single phase symmetry. 
\end{abstract}


One of the central issues 
of solitary wave theory is the question of stability \cite{Pego,Kusmartsev}.
Stability of solitary waves ('solitons') in nonintegrable Hamiltonian  
models is often governed by the  
derivative of a certain invariant with respect to an associated 
parameter of the solution \cite{Pego,Kusmartsev,Vakhitov}. 
For instance, positivity of the derivative of the total energy
with respect to the nonlinearity-induced wave-number shift is
sufficient for stability of bright solitary waves in such
fundamental  processes as
self-action of radiation in media with intensity dependent
refractive index and in degenerate three-wave
mixing in  quadratic nonlinear media \cite{Vakhitov}.

A wide range of parametric processes conserve not only the total
energy of the interacting waves but also other energy invariants. 
These can be, for instance,
energies of individual waves  or energy unbalance of two waves.
Typical examples of such processes are non-degenerate three-wave  
\cite{Karamzin,Stegeman,Kivshar,Peschel97}  
and four-wave mixings \cite{Lundquist98}.
By Noether's theorem every integral of the motion in Hamiltonian models
originates from a corresponding symmetry property. Energy invariants
are often associated with phase (gauge) symmetries.
Although many issues of the dynamics of the multi-wave solitons in models with 
several symmetries still remain to be
understood, their stability threshold in many situations
is given by a zero of the determinant of the Jacobi matrix 
$||\partial_{\kappa_j}Q_i||$ \cite{Kusmartsev,Kivshar,Peschel97}.
Here $Q_i$ and $\kappa_j$ are respectively
the integrals of motion and their associated parameters.

Soliton dynamics in a space where the soliton is
localised in some dimensions but extended in one or more others,
e.g. consideration of a soliton stripe, 
raises the problem of modulational instability (MI) 
along the extra dimensions \cite{Pego}. 
MI of solitary waves has been studied in many fields, 
including plasma physics \cite{Pego}, 
fluid dynamics \cite{Pego,fluids} and optics 
\cite{Pego,Kanashov81,DeRossi97,optcommun,Fuerst97,photo}.
For instance in optics the above-mentioned extra dimensions might be associated
with diffraction and/or group velocity dispersion (GVD). 
An analytical approach to the low-frequency limit
of MI for bright solitary waves was originally developed 
by Zakharov and Rubenchik \cite{Zakharov74a}
for the generalised NLS equation, which has a single phase symmetry. 
This approach is based on asymptotic
expansion near neutrally stable eigenmodes (Goldstone modes)
of the solitary wave excitation spectrum. The presence of such neutral modes 
is directly linked to the symmetries of the model.  It follows that 
systems with a single phase symmetry should be qualitatively 
similar to the NLS case \cite{Zakharov74a}, where the soliton 
{\em always} shows MI in the extra dimension.  The unstable mode is of 
neck type (the soliton stripe breaks into a chain of spots) or snake 
type (the stripe distorts in zig-zag fashion) depending on the relative 
signs of the dispersive terms in the localised and extended dimensions.
All previous studies of solitary wave MI  
\cite{Pego,fluids,Kanashov81,DeRossi97,optcommun,Fuerst97,photo} 
have been restricted to situations with a single phase symmetry, 
and all do indeed exhibit NLS-like MI. For example, in degenerate three-wave 
mixing (3WM), which has a single phase symmetry,
the first analytical results on MI of solitary waves reported by Kanashov 
and Rubenchik \cite{Kanashov81} and recently extended and supported by 
numerical results \cite{DeRossi97,optcommun}, 
typically show the  neck/snake scenario.

The influence of extra phase
 symmetries on MI of solitary waves is still an open issue and is the 
main subject of the present Letter. We will concentrate on {\em non-degenerate} 3WM 
as a typical and practically important example of a solitonic 
model with two phase symmetries, motivated by recent theoretical 
\cite{Kivshar,Peschel97} and experimental \cite{Stegeman,Fuerst97} advances 
in the study of quadratic optical solitons.  

In this Letter we show that MI of solitary waves
in {\em non-degenerate}  3WM 
reveals a new branch of neck-type instability which  can be 
ascribed to the extra phase symmetry and its corresponding neutral mode. 
We show that in media with 
normal GVD this new instability dominates the snake-mode responsible for MI
in the corresponding degenerate 3WM model.
Our theoretical approach and the phenomena predicted by it should extend 
to other  solitonic systems of broad interest with similar symmetry
properties, e.g. to incoherently coupled NLS equations \cite{Manakov}
and non-degenerate 4WM \cite{Lundquist98}.

In non-degenerate 3WM the evolution of suitably 
normalised slowly varying
field envelopes $E_{m}$ ($m=1,2,3$) of three waves 
with carrier frequencies $\omega_m$ ($\omega_1+\omega_2=\omega_3$)
propagating in a dispersive and diffractive quadratic nonlinear medium 
can be modeled \cite{Karamzin,Stegeman,Kivshar,Peschel97} 
by the following system of dimensionless equations: 
\begin{eqnarray}
\nonumber && i\partial_z E_1+\alpha_1\partial_x^2E_1
+\gamma_1\partial_t^2 E_1+E_2^*E_3=0,\\
 && i\partial_z E_2+\alpha_2\partial_x^2E_2
+\gamma_2\partial_t^2 E_2+E_1^*E_3=0,\label{eq1}\\
\nonumber && i\partial_z E_3+\alpha_3\partial_x^2E_3
+\gamma_3\partial_t^2 E_3+E_1 E_2=\beta E_3.
\end{eqnarray}
Transverse $x$, longitudinal $z$ and retarded time $t$ 
coordinates are 
respectively measured in units of a suitable beam width, 
diffraction length and GVD parameter.  The 
$\alpha_m$ and $\gamma_m$  are diffraction and dispersion 
coefficients referred to these scales, and the wave-vector mismatch is 
characterised by $\beta$. We neglect spatial and temporal walk-off effects, 
implicitly assuming that either their spatial and temporal scales are much longer 
than those associated with diffraction and GVD, or that walk-off is compensated by 
special techniques, as in recent experiments on temporal solitons in 
degenerate 3WM \cite{Ditrapani}.
 Henceforth we make the experimentally appropriate 
choice $\alpha_{1,2}=2\alpha_3=0.5$. We also assume all $\gamma_m$ 
either negative (normal GVD) or positive (anomalous GVD), 
leaving the case of mixed dispersion for future work. 
For the sake of simplicity we have restricted our model to one transverse 
dimension, as in a planar waveguide. 

Suppressing the time derivatives for the moment, Eqs. (1) have a 
family of non-diffracting solitonic solutions 
$E_m=A_m(x)e^{i(\kappa_mz+\phi_m)}$, where the $A_m$ are real,  
$\kappa_{1,2}=\kappa\pm\delta$ are positive parameters,  
$\kappa_3=2\kappa>-\beta$ and $\phi_{1,2}=\varphi\pm\psi$,
 $\phi_3=2\varphi$, where $\varphi$ and $\psi$ are arbitrary real constants.  
In general functions $A_m(x)$ must be 
found numerically or approximated variationally 
\cite{Kivshar,Peschel97}.  The free choice of $\varphi$ and $\psi$ implies 
  two phase (gauge) symmetries, which by Noether's 
theorem leads to two conserved quantities. 
These are the total energy  $Q=Q_1+Q_2+2Q_3$, and energy unbalance $Q_u=Q_1-Q_2$, 
or equivalent combinations of the $Q_m=\int dx|E_m|^2$. 
Note that the degenerate case forces $E_1=E_2$, and thus $\psi=0$, 
which suppresses one phase symmetry.

Our primary aim here is to study temporal MI  
due to GVD of these spatial solitons. 
This is most interesting and important when they are 
{\em spatially} stable. Their stability against purely spatial perturbations 
has been studied \cite{Kivshar,Peschel97}, yielding a stability 
boundary which in our notation is given by 
$\partial_{\delta}Q \partial_{\kappa}Q_{u}=
\partial_{\kappa}Q \partial_{\delta}Q_{u}$.
Spatially stable domain  is in fact almost 
the entire domain of soliton existence,  
excluding only a small range of $\kappa,~\delta$ values with 
$\beta<0$ \cite{Kivshar,Peschel97}. 
Close to this region there is also a small domain of bistability \cite{Kivshar}. 
Therefore the existence and nature of temporal MI is the important 
question for almost all parameter values, and in particular for the 
entire half-space with $\beta\ge 0$.

To study MI due to GVD we seek solutions of Eqs. (1) in the form of 
spatial solitons weakly modulated in time at frequency $\Omega\ge0$:
$E_{m} = (A_{m}(x) + (V_{m}(x,z) + iW_{m}(x,z)) \cos\Omega t) 
e^{i(\kappa_{m}z+\phi_m)}$. Setting 
$V_m=v_m e^{\lambda z}$,  $W_m=w_m e^{\lambda z}$, 
we obtain two eigenvalue problems  
$\hat L_-\hat L_+\vec v=\lambda^2\vec v$ 
and $\hat L_+\hat L_-\vec w=\lambda^2\vec w$,
where $\vec v=(v_1,v_2,v_3)^T$, 
$\vec w=(w_1,w_2,w_3)^T$ and
\begin{eqnarray}
\nonumber \hat L_{\pm}=\left[\begin{array}{ccc} 
\pm\hat L_{1} &   A_3  & \pm A_2 \\
A_3 & \pm\hat L_{2} & \pm A_1\\ 
\pm A_2&\pm A_1&\pm\hat L_{3}
\end{array}\right],\end{eqnarray}   
$\hat L_{m}=\alpha_m\partial_x^2-\gamma_{m}\Omega^2-\xi_m$, 
where $\xi_{1,2}=\kappa\pm\delta$ and 
$\xi_{3}=2\kappa+\beta$.  Note that $\hat L_{\pm}$ are 
self-adjoint,  so 
$\hat L_-\hat L_+$ and $\hat L_+\hat L_-$ are adjoint operators 
with identical spectra.  It is thus enough to consider the spectrum 
of one of these operators, e.g. $\hat L_+\hat L_-$.  
We are particularly interested in the discrete spectrum, with
eigenfunctions exponentially decaying at $x\to\pm\infty$.

In general the stability problem can only be solved numerically, 
but for small absolute values of $\lambda$ we can obtain some 
analytical results. Our two phase symmetries, plus the Galilean one,  
generate three neutrally 
stable $(\lambda=0)$ eigenmodes of $\hat L_+\hat L_-$ at $\Omega=0$. These 
 are:
$\vec w_{\varphi}=(A_1,A_2,2A_3)^T$, $\vec w_{\psi}=(A_1,-A_2,0)^T$,
$\vec w_{x}=(xA_1,xA_2,2xA_3)^T$. Assuming that $\Omega\ll 1$ 
we can express $\vec w$ as a 
linear combination of $\vec w_{\varphi}$, $\vec w_{\psi}$, $\vec w_x$
and then use an asymptotic approach to find eigenvectors and
corresponding eigenvalues. This approach can be
applied only if all other eigenmodes 
have eigenvalues obeying the condition $|\lambda|> \Omega$.
In practice this only excludes a small neighborhood of 
the spatial stability boundary discussed above.

Three eigenvalue pairs $\pm\lambda$ are obtained from solvability 
conditions of the first order problems.  One, associated with the 
asymmetric eigenvector 
$\vec w_x$, obeys 
\begin{equation}
\lambda_{x}^2\simeq-\frac{2\Omega^2}{Q}\int 
dx\sum_{m=1}^{3}\gamma_m(\partial_x A_m)^2.
\end{equation}
Clearly the asymmetric mode is  unstable for {\em normal} GVD, which 
corresponds to the snake instabilities found in
NLS \cite{Zakharov74a,fluids} and degenerate 3WM  
\cite{Kanashov81,DeRossi97} models.

The other two eigenvalue pairs are associated with linear combinations 
of the spatially symmetric vectors 
$C_{\varphi}\vec w_{\varphi}+C_{\psi}\vec w_{\psi}$, and 
thus with neck-type instabilities. They are the roots of 
\begin{equation}
a\lambda^4+b\Omega^2\lambda^2+c\Omega^4=0,
\end{equation} where
$a=(\partial_{\delta}Q \partial_{\kappa}Q_{u}-
\partial_{\kappa}Q \partial_{\delta}Q_{u})/2$, 
$b=\partial_{\kappa}Q(\gamma_1Q_1+\gamma_2Q_2)+\partial_{\delta}Q_u
(\gamma_1Q_1+\gamma_2Q_2+4\gamma_3Q_3)+
(\partial_{\kappa}Q_u+\partial_{\delta}Q)(\gamma_2Q_2-\gamma_1Q_1)$, 
and $c = -8(\gamma_1\gamma_2 Q_1Q_2 + \gamma_2\gamma_3 
Q_2Q_3 + \gamma_1\gamma_3 Q_1Q_3)$.
These expressions are quite complicated, 
but yield some important general results.  Clearly $c$ is negative when
all $\gamma_m$ of the same sign. Since 
$a>0$ throughout the spatially monostable domain, it follows that the two 
roots $\lambda^2$ are always real and of opposite sign, so that 
there is {\em always} an unstable neck-type mode. 
Thus we establish {\em coexistence and competition of neck and snake instabilities
for normal GVD}. This is a novel feature 
of the present model, quite different from  previous 
analytical results for NLS \cite{Zakharov74a,fluids} and degenerate 
3WM \cite{Kanashov81,DeRossi97}, where the  snake instability is the 
only one for normal GVD.
This is because only one symmetric neutral mode exists in models with a
single phase symmetry, and it generates 
instability only for anomalous GVD. 
In non-degenerate 3WM there are two neck modes, 
 one stable for normal GVD and unstable for 
anomalous GVD, and {\em vice versa}, since $b$ is odd in the $\gamma_m$.

Simple analytic expressions for growth rates of these neck modes can be obtained in 
several special cases, e.g. in the case of second harmonic generation
($\omega_{1}=\omega_2$).  Setting
$\gamma_{1}=\gamma_2$ and $\delta=0$ (readily achieved in experiment
\cite{Stegeman})
the two eigenmodes have either $C_{\varphi}=0$ or $C_{\psi}=0$, with 
eigenvalues
\begin{equation}
\lambda_{\psi}^2\simeq 2\gamma_1\Omega^2 \frac{Q_1}{\partial_{\delta}Q_1},~~~~
\lambda_{\varphi}^2\simeq 
\frac{4\Omega^2 }{\partial_{\kappa}Q}(\gamma_1Q_1+2\gamma_3Q_3).
\end{equation} 
For $\delta=0$, $\partial_{\kappa}Q$ is positive 
and  $\partial_{\delta}Q_1$ is  negative  
in the spatially monostable domain.
Thus the novel neck instability for $\gamma_1<0$ can be directly attributed 
to the gauge symmetry in the differential phase $\psi$ 
and its associated neutral mode  $\vec w_{\psi}$.
On the other hand $\lambda_{\varphi}$ is associated  
with the usual neck MI for anomalous GVD
in models with a single gauge symmetry 
\cite{Zakharov74a,fluids,Kanashov81,DeRossi97,optcommun}.
The expression for $\lambda_{\psi}^2$ holds also (when $Q_u=0$)
for other solitonic models with
a differential phase symmetry. Note, however, that $\partial_{\delta}Q_1$
can generally have either sign, leading to  instability 
with either normal or anomalous GVD.

Solving the eigenvalue problem numerically, we find that 
in low-frequency limit the instability growth rates 
precisely match those predicted by our perturbation theory, see Fig. 1(a), (b).  As 
$\Omega$ is increased each MI gain curve reaches a maximum and then decreases.   
A typical example of the maximal MI growth rate vs $\beta$ is presented
in Fig. 1(c). Similar plots for $Q_u\ne 0$ and across wide range of 
$\gamma_m$ values show the same behaviour \cite{comment}. 
Thus we conclude that {\em for normal GVD the new
neck instability strongly dominates the snake one}. Note that its
growth rate is maximised, as Fig. 1(a), (b) illustrate, for $Q_u=0$.
For normal GVD the unstable eigenfunctions become weakly confined and
develop  oscillating tails as $\Omega$ increases. Because this 
increases computer demand, we have
plotted growth rates in Fig. 1(a) only for $\Omega$ values 
corresponding to well-localised eigenmodes.  Physically, broader 
eigenmodes have weaker overlap with the soliton, and hence lesser gain.

Spatial profiles of the symmetric eigenfunctions at maximum gain
($\Omega=\Omega_{max}$) are presented in Fig. 2. Despite this being 
well beyond the perturbative limit in which expressions (4) apply, 
the novel neck MI eigenmode still has qualitatively 
the same form as $\vec w_{\psi}$,  
i.e. $w_1=-w_2$, $w_3=0$, indicating that the $\psi$ phase
symmetry  underlies the instability through the whole range of
$\Omega$. Similarly, the  ustable neck mode 
for anomalous GVD is evidently associated with the $\varphi$ symmetry.

To test our linear stability analysis and study the nonlinear
evolution we performed an extensive series of computer 
simulations of the  system (1) with initial conditions in form of a soliton 
stripe perturbed  by spatio-temporal white noise of  order $1\%$. 
Typical simulation results are presented in Fig. 3 and they fully 
support our predictions. 
We chose the size of the computational 
window in the time domain to be $18\pi/\Omega_{max}$, 
and the initial soliton 
stripe rapidly develops nine humps, in accord with the stability 
analysis. During further evolution the modulated stripe forms into 
a train of pulses which either spread (normal GVD) or form 
a persistent chain of three-wave optical bullets (anomalous GVD). 
Due to the initial noise, modes from a band of frequencies close to 
$\Omega_{max}$  are able to grow and compete, and hence the modulations 
in Fig. 3 are somewhat irregular.

A striking difference between Figs. 3(a,c) is that the initially 
imposed translational symmetry of the solitary stripe along the time
dimension is broken in different ways.  
For normal GVD interleaved intensity peaks of 
$E_1$ and $E_2$ are formed, while for anomalous GVD the intensity peaks coincide. 
(Each amplitude is modulated with period $\simeq 2\pi/\Omega_{max}$.) 

This difference is directly related with the spatial form of the 
most unstable eigenvectors. In the case of normal GVD $w_1$ and $w_2$ are out of
phase and $w_3=0$, see Fig. 2(a), leading to the interleaving.
Since  $E_1E_2$ drives $E_3$, the intensity profile of the 
second harmonic becomes modulated with period
$\pi/\Omega_{max}$, see Fig. 3(a$_3$).
Because the overlap of the three fields is diminished by this 
evolution, mutual trapping becomes impossible and the whole 
structure eventually spreads through diffraction and dispersion, 
see Fig. 3(b). For anomalous 
GVD, all three components of most unstable eigenvector are in phase,
 see Fig. 2(b), and thus all three intensities become modulated with the 
same temporal period, see Fig. 3 (c$_1$), (c$_2$), (c$_3$). This provides 
conditions for mutual self-trapping of the filaments, see Fig. 3(d).

The predicted instabilities can be experimentally 
observed in a nonlinear material longer 
than the MI gain length $l_g\sim\lambda^{-1}$ and with pulse width order of several
$2\pi/\Omega_{max}$ or more. Following Ref. \cite{DeRossi97}, 
typical values for KTP $l_g\sim 1cm$ and $2\pi/\Omega_{max}\sim 10^{-12} s$. 
Fig. 1(c) shows that the new neck instability has 
the highest gain and is thus most easily observable. 
Artificially birefrigent semiconductor materials, which are  highly nonlinear, seem 
quite promising, based on a recent  experiments on 3WM \cite{Fiore98}. Waveguides 
containing Bragg structures might be very suitable for 
observation of the MI phenomena predicted here because of their 
large and controllable dispersion \cite{He97}.

In summary,  we have analysed and described dispersive MI 
 of  spatial solitons due to non-degenerate 3WM. 
Using this model as a typical example we generalised a previous analytical approach 
to MI \cite{Zakharov74a} to models with two phase symmetries.
We found that the extra neutral mode associated with the additional 
phase symmetry gives rise to a new branch of MI. 
This is symmetric (of neck type), and is found to dominate the 
asymmetric (snake) instability which is the only MI for normal GVD 
in systems possessing just one phase symmetry.
This result enables a new understanding of the dynamics of 
multi-component solitary waves in terms of their phase symmetry 
properties.  The MI phenomena which we have described for 3WM are likely to 
be generic in other solitonic and nonlinear wave models with two
phase symmetries.

This work was partially supported by EPSRC grant GR/L 27916.

\begin{figure}
\centerline{ \epsfxsize=8cm  \epsffile{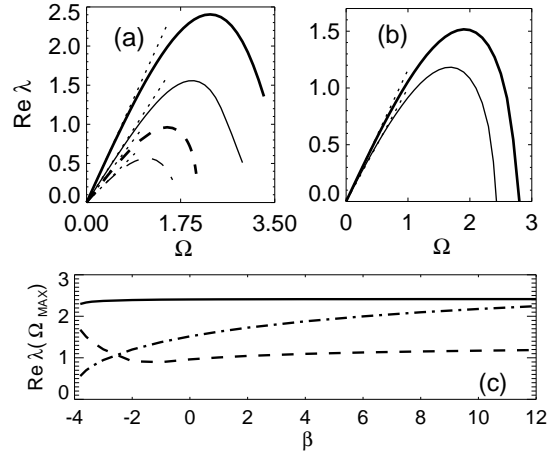}}
\caption{
(a-b) Instability growth rate vs $\Omega$: $Q=65$, $\beta=0$.
Thick (thin) lines are for $\kappa=2,~\delta=0$, $Q_u=0$ 
($\kappa=2.075$, $\delta=1.525$, $Q_u=-36$).  Full (dashed) lines 
correspond to neck (snake) MI. Dotted lines 
are perturbative results. (a) Normal dispersion: 
$\gamma_{1,2}=2 \gamma_3=-0.5$.   (b) Anomalous dispersion: 
$\gamma_{1,2}=2\gamma_3=0.5$. (c) MI growth rate
at $\Omega=\Omega_{max}$ vs $\beta$ for $\kappa=2,~\delta=0$.
Full (dashed) lines correspond to neck (snake) MI 
for $\gamma_{1,2}=2 \gamma_3=-0.5$; dot-dashed line
to neck MI for $\gamma_{1,2}=2 \gamma_3=0.5$}
\end{figure}

\begin{figure}
\centerline{ \epsfxsize=8cm  \epsffile{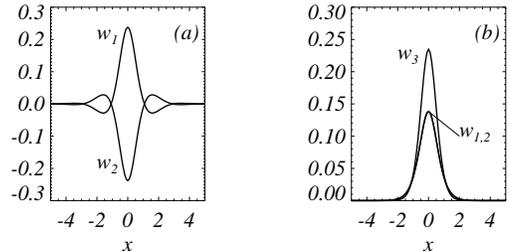}}
\caption{Most unstable eigenmodes ($\Omega=\Omega_{max}$) for $\beta=\delta=0$,
 $\kappa=2$.
 (a) Normal dispersion: $\gamma_{1,2}=2 \gamma_3=-0.5$;
   (b) Anomalous dispersion: $\gamma_{1,2}=2\gamma_3=0.5$.}
\end{figure}

\begin{figure}
\centerline{ \epsfxsize=8cm  \epsffile{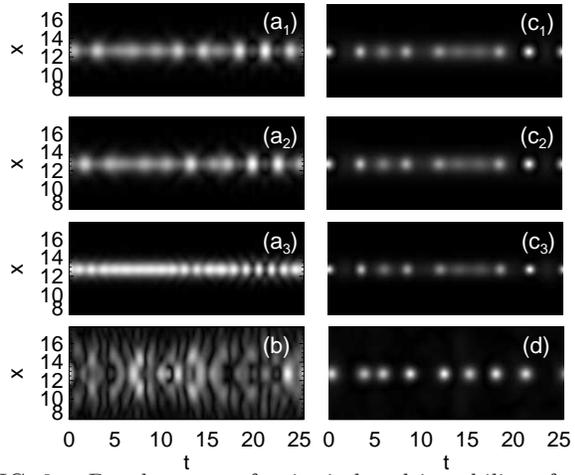}}
\caption{
Development of noise-induced instability of spatial soliton 
stripe: $\kappa=2$, $\delta=\beta=0$. Left (right) panels
 for $\gamma_{1,2}= 2\gamma_3= -0.5$ ($\gamma_{1,2}= 
2\gamma_3=0.5$). (a$_m$) $|E_m|$ at $z=2.7$, (b) $|E_1|$ 
at $z=4.5$, (c$_m$) $|E_m|$ at $z=5.4$, (d) $|E_1|$ at $z=10.8$.}
\end{figure}


\begin{references}
\bibitem{comment}
The expressions for $\lambda_x^2$ and
$\lambda_{\psi}^2$ show that if $-\gamma_3$ is large enough the snake MI has 
the higher gain. However these expressions are only valid in the low-frequency limit, 
and numerics show that for $\Omega=\Omega_{max}$ the neck
MI dominates the snake even more strongly than in the examples shown in Fig. 1.
\end{references}
\end{document}